# Mobile Cloud Computing: A Review on Smartphone Augmentation Approaches[*]

SAEID ABOLFAZLI[1], ZOHREH SANAEI[2], ABDULLAH GANI[3]
Mobile Cloud Computing Research Lab, Faculty of Computer science and Information Technology
University of Malaya
MALAYSIA
abolfazli[1],sanaei[2]@ieee.org, abdullah@um.edu.my[3]

*Abstract:* Smartphones have recently gained significant popularity in heavy mobile processing while users are increasing their expectations toward rich computing experience. However, resource limitations and current mobile computing advancements hinder this vision. Therefore, resource-intensive application execution remains a challenging task in mobile computing that necessitates device augmentation. In this article, smartphone augmentation approaches are reviewed and classified in two main groups, namely hardware and software. Generating high-end hardware is a subset of hardware augmentation approaches, whereas conserving local resource and reducing resource requirements approaches are grouped under software augmentation methods. Our study advocates that consreving smartphones' native resources, which is mainly done via task offloading, is more appropriate for already-developed applications than new ones because of costly re-development process. Moreover, cloud computing is one of the major cornerstone technologies in augmenting computing capabilities of smartphones. Sample execution model for intensive mobile applications is presented and taxonomy of augmentation approaches is devised. For better comprehension, the results of this study are summarized in a table.

*Key–Words:* Mobile Cloud Computing, Smartphone Augmentations, Next Generation Mobile Applications.

## 1 Introduction

Smartphones popularity is drastically increasing and statistics show their surpassing trend toward stationary computing devices [1]. Mobile users are visioning to perform heavy computing tasks as convenient as stationary devices while on the go [2]. However, device handiness and compactness, mobility, and current technologies restrain smartphones to limited computing capabilities due to limitations of underlying resources like CPU, battery, and local storage.

Manufacturers are enhancing computing capabilities of smartphones by generating high-end hardware, but manufacturing costs and sharp rise in device price on one hand and limitations on size, weight, and battery on the other hand hinder their attempts.

Considering the fact that energy, as the essential resource, is the only unrestorable resource in smartphones, several proposals [3, 4] are aimed to not only conserve local energy, but also enable resource-intensive application execution.

Reducing resource requirements of mobile applications is the recent approach to empower mobile devices. The main notion in this approach is to not perform heavy local transactions. Instead, intensive components' execution will be performed in collaborative environment, where powerful computing devices perform code execution on behalf of the weak smartphone. By rapid proliferation of cloud computing technology [5], cloud resources are highly utilized in recent approaches [6,25,36,37]. Cloud computing is envisioned to deliver elastic computing capability to end-user based on "pay as you use" principal and Service-Level Agreement (SLA). SLA is a negotiable part of the service contract, between cloud provider and consumer, to define and enforce service level against an agreed cost. Successful exploitation of cloud resources in stationary machines, motivates elastic resources for mobile devices specially smartphones in wireless ecosystem that breed new computing field, called Mobile Cloud Computing (MCC). MCC deemed to reduce the ownership and maintenance cost of mobile devices and conserve local resources by leveraging cloud distributed resources.

Moreover, MCC is potentially useful to enhance data safety by storing them in the cloud to reduce risk of device malfunctioning and robbery. The recent R&D trends advocate that cloud computing is preferable computing environment for process execution and data storage.

---

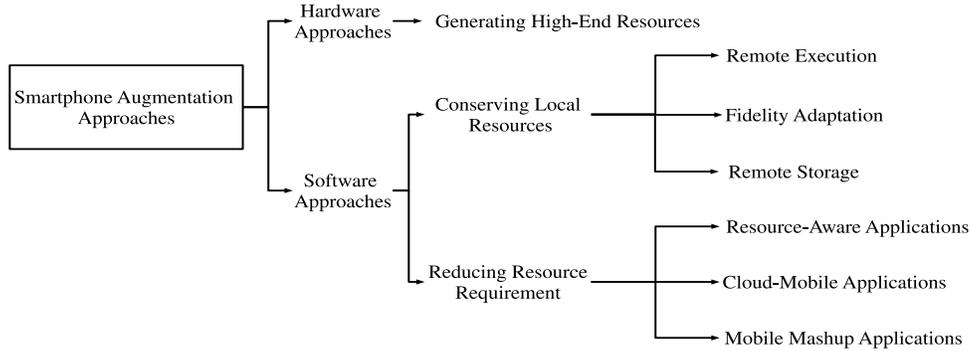

Figure 1: Taxonomy of Smartphone Augmentation Approaches

In this paper, we review augmentation approaches and present taxonomy of augmentation approaches in Section 2. The taxonomy contributes toward advancement of research and development in augmenting computing abilities of smartphones to deliver a rich experience to mobile users. Based on the current augmentation approaches, a sample execution flow of intensive mobile applications is suggested and the paper is concluded in Section 3.

## 2 Taxonomy

Based on available augmentation efforts in literature, this section presents our proposed taxonomy. We have classified augmentation approaches into two main categories, namely hardware and software. In first approach, augmentation is achieved by manufacturing high-end resources and hardware. In second approach, smartphones are augmented either by conserving their local resources, or reducing resource requirements of mobile application. Figure 2 depicts the devised taxonomy and various approaches with relative efforts are described as follows. For better comprehension, the results of this comparison are summarized in Table 1.

### 2.1 Hardware Augmentation

Hardware augmentation is achieved by manufacturing long lasting and high-end hardware.

**Generating High-End Resources:** The initial step in enhancing computing capabilities of smartphones is to increase device's local resources by manufacturing high-end hardware like multicore processor and long lasting battery [7, 8]. However, it is very slow and costly advancement considering current technologies.

Energy is the only unrestorable resource in mobile devices which cannot be renewed without external resource [9]. Current technologies are able to only increase battery capacity 5% per annum [10] while heat dispensation is yet an issue [11]. Many energy harvesting efforts are in progress since 90's to replenish energy from resources like human movement [12], solar energy [13], and wireless radiation [14] which could not noticeably enhance energy deficiency and are still under investigation. However, mobile manufacturers such as Samsung [8] are endeavouring to produce energy-efficient hardware resources with enhance performance.

Similarly, generating large storage and screen decreases smartphone handiness due to additional size and weight. Data storage and retrieval is proven to be energy-hungry[15, 16]. Hence, memory enlargement contributes toward faster battery drainage. In efforts to extend data-presentation area, Kyocera leverages two 4.3" screens to stretch smartphone screen size to 4.7", but it sharply drains the battery (http://www.echobykyocera.com/). Nirvana project [17], GeoTV [18], and remote display solution [19] are some of the efforts to extend smartphones' screen size.

However, manufacturing sophisticated hardware noticeably increases the price of smartphones compare to stationary machines. Moreover, unlike PCs, smartphone's hardware is not upgradable, hence, a new device should be possessed in case of technology advancement. Therefore, alternative techniques are essential to augment smartphones without ownership price hike.

### 2.2 Software Augmentation

The alternative augmentation approach to enhance computing capabilities of smartphones without price hike is achievable using soft techniques that are described as follows.

### 2.2.1 Conserving Local Resources

The early mobile augmentation attempt refers back to late 90' when mobile resources were aimed to be conserved through execution workload reduction. In this domain, noticeable R&D efforts have been taken place which are briefed below.

**Remote Execution:** Rudenko et al, [3] in 1998 introduce the idea of remote execution to conserve mobile device's energy. Later in 2001, Satyanarayana [4] reintroduces the concept of remote execution in pervasive environment and coined the term Cyber Foraging (CF) to not only conserve local resource like battery, memory, and storage of mobile devices, but also enable execution of intense processing applications in smartphone [6, 22]. Thenceforth, a noticeable number of research and development appeared in literature [20-22]

CF is an augmentation method to migrate the resource-hungry components of mobile applications to nearby resource-rich computing machines, called surrogates. Surrogates are connected to uninterruptable power source and the Internet to provide unpaid computing resources. However, in one hand processing overhead due to pre and post offloading tasks (like surrogate discovery, resource estimation, code partitioning, and networking) and in the other hand lack of supervising authority hinder usability of CF methods. In some cases the offloading overhead exceeds conserved resources [23]. Moreover, developing CF-enabled mobile applications creates further programming complexity and cost. Although CF consumes local resources, it seems to be the common approach for enabling execution of the already-developed mobile applications since the re-development process is costly and time consuming.

Moreover, guarantying security of surrogate machines to end-user from one side and convincing surrogate machine's owner that user will not violate terms of service from the other side create hindrances to remote execution. Hence, evolutionary developments of CF methods deemed to mitigate offloading overhead and communication latency in trustworthy environment.

In latest efforts [22, 24, 25] cloud computing is deployed as a resourceful environment to address current problems since security, quality of service, reliability, and expenses are concerned.

**Fidelity Adaptation (FA):** Another potentially useful approach to conserve mobile's resources is fidelity adaptation (FA). This is an alternative solution to augment smartphones in absence of surrogates via exchangeing execution quality for energy conservation [26]. Energy is conserved and application can be executed by altering application-level quality metrics to exchange quality of service for local resources [27]. For example, in multiple-power-state memory [28] quality of service is exchanged with energy consumption [28]. Several approaches [12, 29] leverage composition of CF and FA for better performance.

**Remote Storage:** In capacity-aware data storage approaches, leveraging de-duplication (omitting redundant data) methods and hierarchical remote storages can conserve local storage of smartphone if non-crucial and least frequently used data can be stored remotely. Depending on data type and user requirements (e.g. security, reliability, safety) tiering process ensures desired information availability in the right place at the proper time, assuming availability of reliable connectivity [30]. Example of remote storage is Cloud Storage [31] where not only smartphone storage is virtually expanded, but also data safety is enhanced [32]. However, data transmission, compression, and encryption consume local resources of mobile device.

### 2.2.2 Reducing Resource Requirements

Parallel to manufacturing sumptuous high-end hardware, reducing local resource requirement plays a vital role in augmenting smartphones' processing capabilities. This approach demands consideration in design phase of development with focus on developing resource-efficient applications.

**Resource-aware Algorithm:** Mobile domain is a heterogeneous environment with a plethora of technologies. In wireless communication, there are various non-uniform technologies with different quality promises and energy consumption levels. Considering limited resources of mobile devices, creating resource-aware mobile application promotes considerable resources reduction. For instance, energy-aware applications are required to exploit 2G technology for telephony and 3G for FTP service because of difference between energy requirement of these technologies [15].

Memory intensive applications are also energy-hungry; increase in memory size leads to more power consumption [15, 28]. There are several energy-aware memory management studies [33, 34] to reduce energy consumption of data storage process. Mobile Ram and Phase Change Memory (PCM) [35] are common memory technologies used in smartphones. In Mobile Ram, power management unit maintains multiple power states like 'Self Refresh' and 'Power Down' to minimize power consumption while PCM leverages three states of 'I/O', 'on', and 'off' to stimulate energy efficient data storage. In 'off' mode, cell energy consumption is 0mW, whereas 'on' state consumes 74mW, which is more energy efficient approach compare to Mobile-RAM [28].

Table 1: Comparison of Mobile Augmentation Approaches

| Metrics | Local Resource Consumption | Implementation Cost | Implementation Complexity | Device side Maintenance | Quality of Experience | Network Delay | Execution Delay | Security | Data Safety |
|---|---|---|---|---|---|---|---|---|---|
| Generating High-End Hardware | NA | High | High | Low | High | Low | Low | High | Low |
| Remote Execution | Low | Med | Med | High | Med | High | Low | Low | NA |
| Fidelity Adaptation | Low | Low | Low | Med | Low | Variable | Variable | Variable | Low |
| Remote Storage | Med | Med | Med | Med | Med | High | NA | NA | High |
| Resource Aware Algorithm | Low | Med | Med | Low | Med | Low | Med | Med | Low |
| Mobile Cloud Application | Low | Med | Med | Low | High | Med | Low | Med | High |
| Mobile Mashup Application | Low | Med | Med | Low | High | Med | Low | Med | High |

**Cloud-mobile Applications:** Cloud-mobile applications are envisioned to minimize smartphones' resource consumption by leveraging rich cloud resources with no quality degradation. Cloud components implicitly save smartphone resources. Moreover, large, heterogeneous cloud resources expedite code execution and reduce overall execution time without using native smartphones' resources. Therefore, smartphons monitoring time and communication overhead are shrunk leading to explicit resource saving.

March et al. [36] develop a mobile application framework leveraging cloud resources. In this component-based developing approach, three types of component, namely mobile, cloud, and hybrid (cloud-mobile) are presented to facilitate design and development of mobile applications with least component-level communication and dependency that promises high functionality and resource efficiency. The cloud and mobile components are running on the cloud and mobile device respectively, while hybrid components are free to run either locally or remotely(depends on computing capability of underlying smartphone). In order to reduce resource consumption, resource-intensive instructions are developed as cloud components. During runtime, the execution request is sent to the cloud and results will be sent back to mobile device.

In similar effort, Lu et al. [37] leverage cloud, advanced compression methods, and networking technologies to develop a conceptual architecture to perform screen rendering tasks inside the cloud where the online data and remote code execution take place. Certainly, resource-hungry and non-interactive components of screen will be executed inside the Cloud so that the local CPU, GPU, and battery consumption are reduced.

**Mobile Mashup Application:** Mobile Mashup is another approach that can be leveraged to reduce resource requirement of mobile applications. Mobile Mashup (MM) is a technique to create mobile application by aggregating available *services* and *contents* offered in ubiquitous environment leveraging Mobile Service Oriented Architecture (Mobile SOA) [38] as the backbone. However, from resource efficiency point of view, content mashup is less suitable than functionality, since I/O transactions and content handling are battery-hungry tasks [15].

Considering mobility and dynamism of environment, loose service composition is an essential trait of mashup applications which can be achieved by coarse-grained mashup locally (client side) and fine-grained mashup remotely (server side) [39].

However, mashup by using SOA is a complex approach. Several efforts have been done to justify and fit it into mobile ecosystem and benefit from loosely coupling and remote execution of services toward conserving local resources. Among them, minimizing data transfer between mobile and service provider by coarse grain mashup and shrinking XML file size [40], compressing SOAP message [41] or using RESTful message [30], asynchronous communicating with server [42, 43], and leveraging lightweight workflow [31] are to name few approaches that are aimed to reduce the resource consumption of mobile applications and enable smartphones to delivering rich user experience.

Based on discussion in this paper, a sample mo-bile application execution flow is derived and depicted in Figure 3. Mobile Application execution is highly context-dependent in which several metrics like availability of local resource, user QoS requirement, SLA, and network availability impact at runtime.

## 3  Conclusion

In this paper, we reviewed and synthesized smartphone augmentation approaches from application execution point of view. Generating high-end hardware is expensive, energy-hungry, and time-consuming effort that requires imminent technologies. Smartphones are not intrinsically safe data storage. Data safety and privacy is vulnerable due to risk of device malfunction, robbery, and loss. However, cloud computing is stretched to the mobile domain with aim to reduce necessity of high-end hardware, minimize ownership and maintenance cost in amortized manner, and enhance data safety and privacy.

Conserving local resources through cyber foraging and fidelity adaptation are feasible and widely acceptable approaches to alleviate execution of heavy already-developed applications and adaptive programs on mobile devices. However, several problems challenge the development including security of both user and service providers, adequate estimation of resource consumption, availability of remote resources, maintenance cost, and network delay.

Reducing resource requirements is mainly achieved through novel programming architecture leveraging cloud computing and mashup technologies. Cloud-mobile and mashup applications are flexible distributed applications that request remote execution of resource-intensive parts of applications.

As explained in section 2, mobile cloud computing as the technology of future, provides trustworthy resource-rich environment for mobile users specially smartphone users. Therefore, future generation of mobile applications deemed to be highly dependent to the Cloud services.

**Acknowledgements:** This work is fully funded by Malaysian Ministry of Higher Education under the University of Malaya High Impact Research Grant UM.C/HIR/MOHE/FCSIT/03.

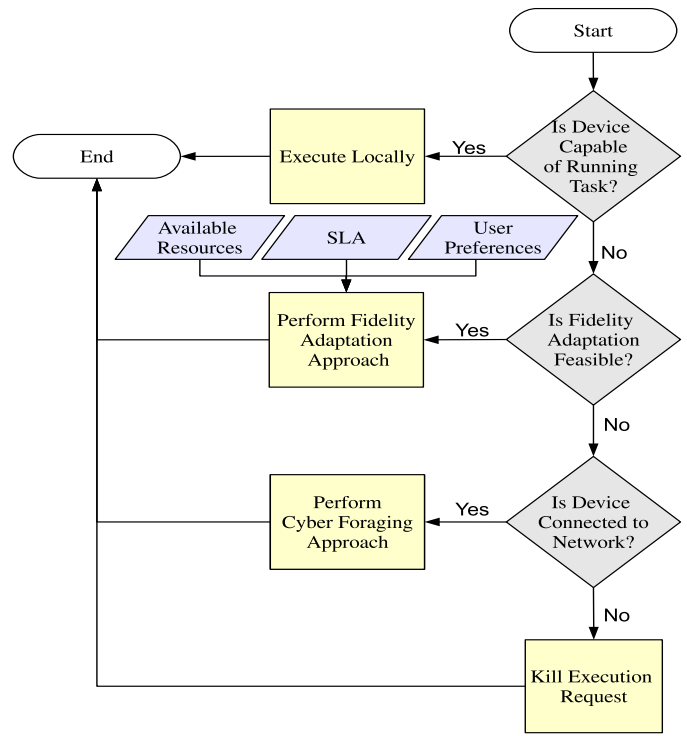

Figure 2: A Sample Mobile Application Execution Flow